\documentstyle[12pt]{article}
\def\lsim{\mathrel{\rlap{\lower4pt\hbox{\hskip1pt$\sim$}}
    \raise1pt\hbox{$<$}}}         
\def\gsim{\mathrel{\rlap{\lower4pt\hbox{\hskip1pt$\sim$}}
    \raise1pt\hbox{$>$}}}         
\def\ut#1{$\underline{\smash{\vphantom{y}\hbox{#1}}}$}

\long\def\caption#1#2{{\setbox1=\hbox{#1\quad}\hbox{\copy1%
\vtop{\advance\hsize by -\wd1 \noindent #2}}}}
\begin{document}
\begin{flushright}
OSU-TA-3/95 \\
DOE/ER/40561-186-INT95-0084
\end{flushright}
\begin{center}
\bf {Implications of the GALLEX Source Experiment \\
for the Solar Neutrino Problem}
\end{center}
\vskip18pt
\begin{center}
Naoya Hata  \\
{\it Department of Physics, Ohio State University, Columbus, Ohio 43210} \\
Wick Haxton \\
{\it Institute for Nuclear Theory, NK-12, \& Department of Physics, FM-15 \\
University of Washington, Seattle, WA  98195}
\end{center}
\begin{abstract}
We argue that, prior to the recent GALLEX $^{51}$Cr source experiment,
the excited state contributions to the $^{71}$Ga
capture cross section for $^{51}$Cr and $^7$Be neutrinos were poorly
constrained, despite forward-angle (p,n) measurements.  We
describe the origin of the uncertainties and estimate their
extent.  We explore the implications of the source
experiment for solar neutrino capture in light of these
uncertainties.  A reanalysis of the $^7$Be and $^8$B flux
constraints and MSW solutions of the solar neutrino puzzle
is presented.
\end{abstract}
\vskip 18pt
Recently the GALLEX collaboration reported the first results
of a $^{51}$Cr neutrino source experiment [1].  The collaboration
stressed the importance of this measurement as a test of experimental
procedures, including the overall recovery efficiency of the
product $^{71}$Ge.  This test is in addition to the run-by-run checks
on the chemical extraction efficiency that have been performed
by introducing Ge carrier.  These have consistently indicated Ge yields
of about 99\% [2].

     The significance of the source experiment
is that it tests the recovery of $^{71}$Ge under production
conditions almost precisely mimicking those of solar neutrinos.
For instance, the recoil energies and atomic excitations accompanying
solar neutrino absorption could conceivably drive a chemical
reaction that would bind  $^{71}$Ge within the detector.  This
possibility is not entirely academic in view of the early
GALLEX experience with cosmogenic $^{68}$Ge ($\tau_{1/2}$ = 270.8 d),
which was not purged as expected from the detector
when the experiment was begun [3].  Rather than continuing
to decline exponentially, the $^{68}$Ge level plateaued after
repeated extractions at a level 20 times higher that the
expected standard solar model (SSM) $^{71}$Ge yield.  While this difficulty was
overcome by heating the tank, it illustrates that the chemistry of reactive
species, when performed at levels below 100 atoms, can be
subtle.

One motivation for the present study is to consider the
role of excited states in $^{71}$Ge in the capture of $^{51}$Cr and $^7$Be
neutrinos.  Electron capture on $^{51}$Cr produces two line sources of
neutrinos of energy 746 keV (90\%) and 431 keV (10\%).  As illustrated
in Fig.~1, the 431 keV neutrinos excite only the ground
state of $^{71}$Ge.  The strength of this transition is fixed
by the known lifetime of $^{71}$Ge, 11.43 days, yielding
\begin {equation}
{\rm  BGT (gs)} = {1 \over 2 J_i + 1} |\langle J_f \|O^{J=1}_{\rm {GT}}\|J_i
\rangle|^2 =
0.087 \pm 0.001
\end {equation}
for the Gamow-Teller (GT) matrix element in the direction
$^{71}$Ga ($J^\pi_i  = 3/2^-$) to $^{71}$Ge ($J^\pi_f  = 1/2^-$).  The GT
operator is
\begin {equation}
O^{J=1}_{\rm {GT}} = \sum^A_{i = 1} \vec \sigma (i) \tau_+ (i).
\end {equation}
However the dominant 746 keV branch can excite not only the
ground state but also two other allowed transitions, to the
5/2$^-$ (175 keV) and 3/2$^-$ (500 keV) states in $^{71}$Ge.  These two
excited states also contribute to the absorption of $^7$Be
solar neutrinos.  The allowed transition strengths have not
been measured directly, though arguments based on nuclear
systematics and analyses of forward-angle (p,n) cross sections
have led to a ``standard" estimate of their contribution to
the $^{51}$Cr experiment of 5\% [4].  The GALLEX collaboration
adopted this estimate in deducing [1]
\begin {equation}
R = (1.04 \pm 0.12)~~~  (1 \sigma),
\end {equation}
where $R$ is the ratio of the measured $^{71}$Ge atoms to expected
in the source experiment.

Let us begin with a restatement of the source experiment
result that is free of nuclear structure assumptions,
\begin {equation}
R_0 \equiv E \left[1 + 0.667 {\rm {BGT} (5/2^-) \over \rm {BGT
(gs)}} +
0.218 {\rm {BGT} (3/2^-) \over \rm {BGT (gs)}} \right] = 1.09 \pm 0.13,
\label {xxx3}
\end {equation}
where BGT(gs), BGT(5/2$^-$), and BGT(3/2$^-$) are the Gamow-Teller
strengths for the ground state and first two excited states in $^{71}$Ge and
$E$ represents any deviation in the overall \ $^{71}$Ge detection
efficiency from that \ calculated and used by \  the experimentalists.   The
coefficients of the second and third terms within the brackets
represent the phase space for exciting the 175 and 500 keV
states by $^{51}$Cr neutrinos, normalized to the ground state phase space.
The experimental quantity $R_0$ is also normalized to the
ground state contribution.  Thus $R_0$ = 1.05 is the ``standard"
value corresponding to a 5\% excited state contribution to the
absorption of $^{51}$Cr neutrinos.

If the excited states comprise only 5\% of the $^{51}$Cr neutrino
capture rate, the transitions to these states must be roughly an order
of magnitude weaker than to the ground state.  Do we know
this with certainty?  We know of three arguments that address
this issue:

1)   Bahcall's original estimate [5] of the strengths of the excited
state matrix elements was made by examining known transitions
in neighboring nuclei that connect states of the same spin and
parity.  In the case of the 3/2$^-$ to 5/2$^-$ transition, eight
such decays were found, and from these Bahcall deduced
BGT(5/2$^-$) $<$ 0.004.  Three or four of these decays, though
not all, appear to be related to the Ga transition when
viewed from the perspective of a single-particle model like that of
Nilsson.

The neutron number of $^{71}$Ga is 40; the analogous neutron
numbers for the comparison transitions range from 33 to 38.
Clearly one needs to assume that the nuclear structure is
reasonably constant under addition of neutrons to use these
data to extrapolate to the N=40 case of interest.

Both empirical and theoretical considerations indicate that this
assumption is unwarranted.  The empirical approach is to test this procedure
in the case of the known $^{71}$Ga $\to ^{71}$Ge(gs) transition, BGT = 0.087.
The most closely analogous neighboring decay is $^{73}$Ga(3/2$^-$) $\to
^{73}$Ge(1/2$^-$, 67 keV): In the Nilsson model this transition
differs only by the addition of two spectator neutrons
to the lowest deformed level based on the 1g$_{9/2}$ spherical shell.
Yet for this transition, BGT = 0.0016, or more than a factor of 50
weaker.

The theoretical considerations come from evidence [6,7]
that the behavior of N $\sim$ 40 Ge isotopes under changes in the
neutron number is spectacularly nonlinear due to the interplay of
coexisting spherical and deformed bands.  Near N=40 it is possible
to produce energetically favored, highly deformed neutron
configurations in which a pair of neutrons occupies a
Nilsson orbital based on the 1g$_{9/2}$ spherical shell.  This
occupancy in turns polarizes and deforms the proton orbitals: the
strong 1g$_{9/2}$(n)$-$1f$_{5/2}$(p) interaction (these shells have similar
nodal
structure) drives protons from the 2p$_{3/2}$ shell into the 1f$_{5/2}$ shell.

     This explanation accounts for an apparent level crossing of
the ground-state and first excited 0$^+$ bands in even-N Ge isotopes
near N=40 (see Fig. 2 of Ref. [8]).  Measurements show a
corresponding sudden change in the shells occupied by the
four valence protons: The ratio of the 2p$_{3/2}$ and 1f$_{5/2}$ spectroscopic
factors plunges from 2.3 to 0.9 when two neutrons are added to $^{72}$Ge
(Fig. 3 of Ref. [8]).
A large-basis shell model calculation has qualitatively reproduced
this behavior, showing rapid changes in the proton
occupancies between N = 38 and 42.

 $^{71}$Ga ($^{71}$Ge) can be naively viewed as a proton
(neutron) hole in a $^{72}$Ge (N=40) core.  As the core changes
dramatically as the neutron number varies, $\beta$ decay
strengths might be expected to evolve sharply between N=38 and 42.
Presumably this accounts for the disparity between the $^{71}$Ga
and $^{73}$Ga BGT values noted above.

Similar arguments apply to the transition to the 3/2$^-$ (500 keV)
state.  We conclude that systematics do not place quantitative
constraints on the matrix elements of interest.

2)   One could appeal to theory, but it is difficult to estimate
weak BGT values reliably.  Two credible attempts have been
made, by Mathews et al. [9] and by Grotz, Klapdor, and Metzinger [10].
Their predictions are BGT(5/2$^-$)/BGT(gs) = 0.23 and 0.001
and BGT(3/2$^-$)/BGT(gs) = 0.014 and 0.86, respectively.  Given
the large discrepancies, they provide little guidance.  Neither
calculation incorporates the physics that we argued in 1)
drives the dramatic shape changes near N=40.  For example, the
polarizing 1f$_{5/2}$(p)$-$1g$_{9/2}$(n) interaction plays no role in the
Mathews et al. calculation since neutrons are not allowed into
the 1g$_{9/2}$ shell.

3)   We thus conclude that (p,n) reactions are the one hope for
quantitatively constraining the unknown BGT values.  At medium energies
the proportionality between forward-angle cross sections and BGT
strength has been well established [11] in the case of strong
transitions (BGT $\gsim$ 0.4).  For this reason (p,n) mappings of
the broad profile of BGT strength have been considered a valuable
tool for estimating $^8$B solar neutrino cross sections.  However,
it will become apparent below that the use of (p,n)
reactions to constrain single transitions with small BGT strengths
is a far more speculative endeavor.

Measurements for $^{71}$Ga were made at 120 and 200 MeV by Krofcheck et
al. [12], yielding
\begin {equation}
\rm {BGT}_{\rm {(p,n)}} (5/2^-) < 0.005  ~~ \rm {and}
{}~~\rm {BGT}_{\rm {(p,n)}} (3/2^-) = 0.011 \pm
0.002.
\label {xxx4}
\end {equation}
Perhaps because they conform to arguments based on systematics,
these results appear to have been accepted rather uncritically.

But as these results are crucial to the interpretation of
the calibration experiment, it is important to try to assess
their likely reliability.  Are (p,n) reactions
a reliable probe of BGT values $\sim$ 0.01?
And if not, what is a reasonable error bar to assign
to these determinations for very weak transitions?

We can try to answer these questions by examining (p,n)
results for transitions of known strength.  Ten transitions [13] for
the 1p and 2s1d shell are shown in Table 1, including five
mirror transitions where the nuclear structure is likely
quite simple.  In five cases the deduced (p,n) BGT values are quite
large ($\gsim$ 1.0), three are somewhat weaker ($\lsim$ 0.5), and two
are very weak ($\sim$ 0.01, comparable to the Krofcheck $^{71}$Ga excited
state BGTs).  The correspondence between the (p,n) and $\beta$
decay results for the five strong transitions is typically 10\%,
with one case showing a 30\% discrepancy.  The proportionality
for the three weaker transitions is rather poor (typically off
by a factor of two).  The discrepancies for the two very weak
transitions are very large, as the (p,n) BGT values exceed the $\beta$
decay values by factors of $\sim$ 7 and $\sim$ 100.

The studies of Refs. [14] and [15]  both identified an
(L=2 S=1)J=1 term in the (p,n) operator as a likely source of
these discrepancies.  This operator arises in distorted wave
Born treatments of (p,n) scattering and has been shown to
affect weaker transitions in $^{37}$Cl(p,n)$^{37}$Ar substantially
(e.g., BGT $\lsim$ 0.1) [14].  Watson et al. [15] attributed the
discrepancies in Table 1 to the effective tensor operator
$O^{J=1}_{L=2}$
\begin {equation}
\label {xxx17}
\langle J_f \|O^{J=1}_{\rm {(p,n)}} \|J_i \rangle = \langle J_f
\|O^{J = 1}_{\rm {GT}}
\|J_i
\rangle + \delta \langle J_f \| O^{J = 1}_{L=2} \| J_i \rangle_{\rm {SM}}
\end {equation}
where
\begin {equation}
O^{J = 1}_{L=2} = \sqrt{8 \pi} \sum^A_{i = 1} [Y_2 (\Omega_i) \otimes
\vec \sigma
(i)]_{J = 1} \tau_+ (i) .
\label {xxx5}
\end {equation}
In these studies values of the tensor operator coefficient $\delta$ = 0.073
(0.064)
were chosen for the 2s1d (1p) shell.
The parameterization used in [15] would give
$\delta$ = 0.097 for $^{71}$Ge.  The notation $\langle \|~~~\|
\rangle_{\rm {SM}}$
indicates that a shell model reduced matrix element is to be taken.

The operator $O^{J=1}_{\rm {(p,n)}}$ defines an effective (p,n) BGT value,
BGT$^{\rm {eff}}_{\rm {(p,n)}}$.  The calculated values are given in
Table 1.  The shell model matrix elements of $O^{J=1}_{L=2}$ were
evaluated using Cohen and Kurath [16] (1p shell) and Brown-Wildenthal [17]
(2s1d shell)
wave functions.  We also take the sign of the interference between
$O^{J=1}_{\rm {GT}}$ and $O^{J=1}_{L=2}$ from the shell model.
The magnitude of $\langle J_f \|O^{J=1}_{\rm {GT}} \|J_i \rangle$ was
taken from the $\beta$ decay BGT values.  The results match the measured
BGT$_{\rm {(p,n)}}$ rather well: Phenomenologically the differences
between $\beta$ decay and (p,n) BGT values do appear to be consistent
with an additional L=2 tensor operator contributing to the
latter.  We determined $\delta$ by a
least squares fit to the measured values, yielding $\delta$ = 0.069 (0.096)
for the 2s1d (1p) shells, results reasonably close to those
recommended in Ref [15].  (Our treatment differs slightly from that of
Ref. [15] because we express the difference in the (p,n) and
$\beta$ decay matrix elements as a shell model matrix element
of an effective operator, rather than introducing effective
operators for both (p,n) reactions and $\beta$ decay.
The resulting values for $\delta$ differ from those of Ref. [15] for this
reason and because some of the $\beta$ decay BGT values of Table 1
have been updated.)

This ansatz provides some insight into possible uncertainties
in (p,n) BGT mappings.
It is apparent, for very suppressed GT transitions, that (p,n) BGT
values $\sim \delta^2 \sim$ 0.01 can arise solely from the
L = 2 contribution to Eq. (\ref {xxx17}).  This accounts for the
two most dramatic discrepancies in Table 1.   (And as the (p,n) BGT
values of Eq. (\ref {xxx4}) are $\lsim$ 0.01, it follows
that these transitions could also be characterized by vanishing
GT strengths.)  The remaining significant discrepancies in Table 1
involve the analog transitions in $^{13}$C, $^{15}$N, and
$^{39}$K, which would be described naively as either 1p$_{1/2} \to$ 1p$_{1/2}$
or 1d$_{3/2} \to$ 1d$_{3/2}$ transitions.  But for single-particle transitions
with $\ell = j -1/2$
\begin {equation}
{\delta \langle J_f \|O^{J = 1}_{L = 2} \| J_i \rangle \over \langle J_f \|
O^{J =
1}_{\rm {GT}} \|J_i\rangle} = 2 \delta \left({\ell + 1 \over 2 \ell - 1}
\right).
\end {equation}
Thus the L=2 and GT operators interfere constructively.  This
explains why the (p,n) BGT values of Table 1 are too large.

The case of $^{71}$Ga is quite different, however.  The 1/2$^-$,
5/2$^-$, 3/2$^-$ level ordering in $^{71}$Ge in consistent with a
neutron shell of moderate positive deformation $\beta \sim$ 0.05 $-$ 0.15
in the Nilsson model.  Thus the 5/2$^-$ 175 keV state is likely
associated with a neutron orbital that is almost entirely 1f$_{5/2}$.
We expect the valence protons to occupy the 2p$_{3/2}$ shell,
primarily.  Thus the transition density [18] is likely dominated by
\begin {equation}
1f_{5/2}(n) \to 2p_{3/2}(p),
\nonumber
\end {equation}
an amplitude that does not contribute to the GT operator but
generates the strongest L=2 matrix element in the 1f2p shell.
The competing GT amplitude would arise from presumably less
important terms in the density matrix, e.g., 2p$_{1/2} \to$ 2p$_{3/2}$
and 1f$_{5/2} \to$ 1f$_{5/2}$.  We have no experimental information on
the relative sign of the L=0 and L=2 amplitudes.

The strength of this L=2 transition could quite plausibly approach
the single-particle limit [19].  This provides the bound
\begin {equation}
|\delta \langle 5/2^- \| O^{J = 1}_{L = 2} \| 3/2^-\rangle_{\rm {SM}}|\lsim 6
\delta \sqrt
{3 \over 5} = 0.45.
\end {equation}
Using this constraint in Eqs. (\ref {xxx4}) and (\ref {xxx17}) then yields

\begin {equation}
0 \lsim \rm {BGT}(5/2^-)\lsim 0.087.
\label {xxx8}
\end {equation}
That is, for destructive interference a BGT value on the order
of BGT(gs) is not excluded by the (p,n) measurements.

The transition to the 3/2$^-$ (500 keV) state is more
complicated.  In an effort to avoid exaggerating the BGT range,
we take some guidance from the Nilsson model, which associates
this state with a neutron hole in a K=3/2 orbital whose
spherical parentage is 2p$_{3/2}$, but which crosses and strongly
mixes with a second K=3/2 orbital whose parentage is 1f$_{5/2}$.
We expect
\begin {equation}
\delta \langle 3/2^-\| O^{J = 1}_{L = 2} \|3/2^-\rangle_{\rm {SM}} \sim 6
\delta
\sqrt {3
\over 5} \left[\Psi_{2 p_{3/2}1f_{5/2}} + {2 \over 9} \Psi_{2 p_{3/2} 2
p_{3/2}}\right],
\label {xxx9}
\end {equation}
where $\Psi_{\alpha \beta}$ denotes components of the one-body transition
density
matrix [18].  But $\Psi_{2 p_{3/2} 2 p_{3/2}}$ likely accounts for the largest
contribution to the GT matrix element, too, to which it contributes with the
same sign as above.  Thus cancellation between the the L=2 and
GT matrix elements likely requires cancellation between the
density matrix elements in Eq. (\ref {xxx9}).  We bound the L=2 matrix
element by taking the single-particle limit under the Nilsson
model constraint that $|\Psi_{2 p_{3/2} 1f_{5/2}}|\lsim {1 \over \sqrt 2}$,
\begin {equation}
|\delta \langle 3/2^-\| O^{J = 1}_{L = 2} \|3/2^-\rangle_{\rm {SM}}|\lsim 6
\delta \sqrt {3
\over 5} \left({7 \over 9 \sqrt 2}\right) = 0.25.
\end {equation}
Combining this with the (p,n) BGT value of 0.011 $\pm$  0.002 yields
\begin {equation}
0 \lsim \rm {BGT} (3/2^-) \lsim 0.057.
\label {xxx11}
\end {equation}

In their discussions of the implications of the source
experiment for solar neutrino capture in $^{71}$Ga, the GALLEX
collaboration fixed the $^{51}$Cr excited state contributions
at 5\% and consider only the affects of shifting this strength
between the 5/2$^-$ and 3/2$^-$ states.  We now would like to make
three observations based on the nuclear structure arguments
of this paper.

$i$)   Without the calibration experiment, no convincing argument
exists for more restrictive bounds on BGT(5/2$^-$) and BGT(3/2$^-$)
than those given by Eqs. (\ref {xxx8}) and (\ref {xxx11}), giving the region
enclosed
by the dashed lines in Fig. 2.  These bounds allow
the total excited state contribution to the $^{51}$Cr capture rate
to range between 0 and 80\% of the ground state contribution,
in contrast to the 5\% employed in the GALLEX calibration discussions.

     The other experimental checks [2] performed by the GALLEX collaboration
make it likely that $E \sim$ 1.0.  We make this assumption now in
order to explore the consequences of the nuclear physics uncertainties,
independent of the question of efficiencies.  It then follows that
the pp capture rate is determined by the known value of BGT(gs).  (The error
associated with the 1\% of captures to the first excited state
is insignificant.)   The $^8$B capture rate depends on the
broad profile of BGT strength up to the $^{71}$Ge particle
breakup threshold of 7.42 MeV.  For states above 500 keV,
we take the ``best value" for this profile from Krofcheck et al., but
associate a
1$\sigma$ normalization uncertainty of 25\%.  [The
Krofcheck profile was normalized by the isobaric analog state (IAS)
transition.  Uncertainties associated with this include the
GT strength beneath the IAS peak, and the reliability of the
calculated Fermi/GT strong distortion factor ratio.  The
25\% uncertainty results from empirical tests [11,20] of the IAS/Fermi
proportionality [21]].  Finally, the $^7$Be cross section has the large
uncertainty associated with the freedom in BGT(5/2$^-$)
and BGT(3/2$^-$), parameters for which we lack even best values.

     Figures 3a, 3b, and 3c show the resulting
constraints on $\phi(^7$Be) and $\phi(^8$B) that we
have extracted from the GALLEX and SAGE [22] experiments, from GALLEX/SAGE in
combination with Kamiokande II/III [23], and by considering all experiments
(GALLEX,
SAGE, Kamiokande II/III, and $^{37}$Cl [24]) together.  The $\chi^2$ fits
include
the effects of pep and CNO neutrinos, and all fluxes are constrained
by the condition that the solar luminosity is fixed [25].  The heavily shaded
regions
in these graphs indicate the allowed fluxes when the unknown BGT
values are assigned their maximum values.  These regions expand to
include the lightly shaded portions when no excited state $^7$Be
capture is assumed.  The dashed line is
the result using Krofcheck's BGT values.  These graphs illustrate
that the unknown nuclear physics induces a considerable uncertainty
in the extracted bounds, represented by the lightly shaded regions,
with large values of BGT(5/2$^-$) and BGT(3/2$^-$) leading
to more stringent constraints on $\phi(^7$Be).

$ii$)   If we now continue with the assumption that $E \sim$ 1.0 but use the
results on the source experiment, a constraint is imposed on the
unknown BGT values,
\begin {equation}
\alpha \equiv 0.667 {\rm {BGT}(5/2^-)  \over \rm {BGT (gs)}} +   0.218
{\rm {BGT}(3/2^-)
\over \rm {BGT(gs)}} = 0.09 \pm 0.13 ~~~  (1 \sigma),
\label {xxx12}
\end {equation}
where it is understood that this quantity is positive.  This
constraint significantly reduces the allowed region for the BGT values, as
illustrated in Fig. 2.  But more important, it almost completely
removes the unconstrained nuclear physics uncertainties that
affect the $^7$Be capture rate.  The capture rate can be
reexpressed in terms of the constrained parameter $\alpha$,
\begin {eqnarray}
\langle \sigma \phi (^7{\rm {Be}})\rangle =
(1.3 ~{\rm {SNU}})~ P_{\rm {MSW}}
(384~ {\rm {keV}})
 ~+~   (34.4 {\rm {~SNU}})~ P_{\rm {MSW}} (862~ {\rm {keV}})
\nonumber \\
 \times \left[1 + 1.09 \alpha +
0.080 \left[ -0.218 {\rm {BGT}(5/2^-) \over \rm {BGT(gs)}} +
0.667 {\rm {BGT}(3/2^-) \over \rm {BGT(gs)}} \right]  \right],
\label {xxx13}
\end {eqnarray}
where P$_{\rm {MSW}}$ denotes a possible reduction in the 862 keV and
384 keV $^7$Be line fluxes due to the Mikheyev-Smirnov-Wolfenstein (MSW)
effect [26].  We have used the latest SSM fluxes of Bahcall and Pinsonneault
(BP) [27]
with He and metal diffusion in deriving Eq. (\ref {xxx13})
($\phi$($^7$Be) = 5.15E9/cm$^2$s).
The last term (in brackets) on the right-hand side of Eq. (\ref {xxx13})
represents
the degree of freedom in the BGT(5/2$^-$) $-$ BGT(3/2$^-$) plane
that is orthogonal to $\alpha$, and thus is
unconstrained by the source experiment.  The small coefficient of
this term reflects the fact that the $^7$Be and $^{51}$Cr cross sections depend
on nearly identical linear combinations of BGT(5/2$^-$) and BGT(3/2$^-$).
Consequently,
the residual, experimentally unconstrained nuclear physics uncertainties in
the $^7$Be
cross section make at most a 3\% contribution, given the bounds in
Eqs. (\ref {xxx8}) and (\ref {xxx11}).
Thus, in principle, a perfect $^{51}$Cr source experiment could
determine the $^7$Be cross section to $\pm$ 1.5\%.
Future improvements in the source experiment will continue to
be well motivated until a comparable statistical accuracy is achieved.

Presently, the uncertainty in the experimental constraint (Eq. (\ref {xxx12}))
is considerably larger than this $\pm$ 1.5 \% ``irreducible" error.  However,
as experiment has determined a ``best value" and error for $\alpha$,
this constraint can now be included in the $\chi^2$ fit.
In Figs. 4 we present calculations analogous to those in
Figs. 3, but with the source constraint included.  Because the
unconstrained uncertainties in the $^7$Be cross section have been
reduced to such a modest level, the boundaries of the allowed regions
can now be represented accurately as lines.

The solar neutrino data, now with the gallium $^7$Be cross
section uncertainties clearly under control, are in serious
conflict with suggested astrophysical explanations.  Nonstandard
solar models generally reduce $\phi(^8$B) more than $\phi(^7$Be),
in contradiction to the data (Fig. 4c).  This difficulty persists
when one considers only the SAGE/GALLEX and Kamiokande data
(Fig. 4b), or any other pair of the SAGE/GALLEX, Kamiokande,
and Homestake experiments (see, e.g., Ref [25]).

We can also now include the source experiment in fitting the
results of the SAGE, GALLEX, Kamiokande II/III, and $^{37}$Cl
experiments in the presence of MSW oscillations.  The MSW solutions
provide an excellent description of the data, as shown in Fig. 5.
We have assumed that the oscillation is into muon or tauon neutrinos,
which will contribute to the Kamiokande II/III signal, though with a
cross section about 1/7 that of electron neutrinos.  The BP SSM with
He and metal diffusion has again been used in the calculations.
We have incorporated the theoretical uncertainties and their
correlations, the Earth effect, the Kamiokande day-night data,
and the improved definition of confidence level contours,
following Ref. [28].

$iii$)   We have assumed in our discussions that the relative efficiency
$E$ can be assumed to be unity.  Even without this assumption,
however, the $^7$Be capture rate can be
reexpressed in terms of the experimental quantity $R_0$,
\begin {eqnarray}
\langle \sigma \phi (^7{\rm {Be}})\rangle = E (1.3 ~{\rm {SNU}})~ P_{\rm {MSW}}
(384~ {\rm {keV}})
+ R_0 (34.4 ~{\rm {SNU}})~ P_{\rm {MSW}}~ (862~ {\rm {keV}})
\nonumber \\
\times \left[1 + {0.043~ \rm {BGT} (5/2^-)
+ 0.073~ \rm {BGT} (3/2^-) \over \rm {BGT (gs)} + 0.667~ \rm {BGT} (5/2^-) +
0.218 ~\rm {BGT}
(3/2^-)} \right].
\end {eqnarray}
The remaining nuclear structure uncertainties affecting the capture
of 862 keV neutrinos varies from 1.0 to 1.05, given the constraints
(Eqs. (\ref {xxx8}) and (\ref {xxx11}))
on BGT(5/2$^-$) and BGT(3/2$^-$).  Thus a measurement of $R_0$ with
absolute precision would determine the overall $^{71}$Ga detector
$^7$Be neutrino rate to $\pm$ 2.5\%, independent of any assumptions about $E$.

If one adopts the extreme view that $E$ is unconstrained
apart from Eqs. (\ref {xxx3}), (\ref {xxx8}), and (\ref {xxx11}), the GALLEX
result
\begin {equation}
\langle \sigma \phi\rangle_{^{71}\rm {Ga}} = 79 \pm 10 \pm 6 ~~~\rm {SNU}
{}~~(1
\sigma)
\end {equation}
is not in serious conflict with the SSM: Maximal values
for BGT(5/2$^-$) and BGT(3/2$^-$) allow $E$ to be as low as 0.54,
implying effective pp and $^8$B neutrino counting rates of 38 and 9 SNU
in the BP SSM.  With a $^7$Be neutrino rate of 38 SNU
and pep and CNO cycle neutrino contributions of $\sim$ 7 SNU, the observed
counting rate would be $\sim$ 90 SNU, within 1$\sigma$ of the GALLEX best
value.  We are not advocating such a view, but instead pointing
out the essential role the other checks on the chemical
extraction and $^{71}$Ge counting efficiencies still play in the
gallium experiment.  These checks are not superseded by the
source experiment.

We thank E. Adelberger, S. Austin, N. Anantaraman, G. Bertsch, B. A. Brown,
and especially P. Langacker for helpful discussions.
This work was supported in part by the U.S. Department of Energy
under grants \#DOE-AC02-76-ERO-3071 and \#DE-FG06-90ER40561 and by
NASA under grant \#NAGW2523.

\pagebreak

\caption{Table 1:}{Comparison of experimental $\beta$-decay BGT values,
experimental (p,n) BGT values, and BGT$^{\rm {eff}}_{\rm {(p,n)}}$
calculated from
the effective operator of Eq. (\ref {xxx17}), using $\delta$=0.069 (0.096)
for the
2s1d (1p) shell.}
\vskip 18pt
\hskip2.85in
\ut{~~~~~~~Experiment~~~~~~~~}
$$\vbox {\tabskip 2em plus 3em minus 1em
\halign to \hsize{\hfil #\hfill && #\hfil \hfil \cr
A$_i$&J$_i$
&J$_f$ (E$_f$ (MeV))
&BGT~$^a$
&BGT$_{\rm {(p,n)}}^b$
&BGT$^{\rm {eff}}_{\rm {(p,n)}}$$^c$\cr
\noalign{\hrule}
\noalign {\vskip 12pt}
$^{13}$C&1/2$^-$ &1/2$^-$ (0.0)&0.20&0.39&0.40 \cr
$^{14}$C&$0^+$ &1$^+$ (3.95)&2.81&2.82&2.84\cr
$^{15}$N&1/2$^-$ &1/2$^-$ (0.0)&0.25&0.54&0.53\cr
$^{17}$0&5/2$^+$&5/2$^+$ (0.0) & 1.05 &0.99&1.15\cr
$^{18}$0&0$^+$ &1$^+$ (0.0) &3.06&3.54&3.11\cr
$^{19}$F&1/2$^+$ &1/2$^+$ (0.0)&1.62&2.13&1.65\cr
$^{26}$Mg&0$^+$&1$^+$ (1.06)&1.10&1.14&1.20\cr
$^{32}$S&0$^+$ &1$^+$~ (0.0)&0.0021&0.014$^d$&0.016$^e$\cr
$^{39}$K&3/2$^+$&3/2$^+$ (0.0)&0.27&0.39&0.39\cr
$^{39}$K&3/2$^+$ &1/2$^+$ (2.47)&0.00017&$\sim$~0.017&0.014\cr
\noalign {\vskip 12pt}
}}$$
\noindent
$^a$Deduced from the compilations of Ref. [29]
\vskip12pt
\noindent
$^b$From Ref. [15] unless otherwise noted.
\vskip12pt
\noindent
$^c$Matrix elements of $O_{L=2}^{J=1}$ and the sign of the interference between
$O^{J=1}_{\rm {GT}}$
and $O^{J=1}_{L=2}$ were evaluated with Cohen and Kurath (1p
shell)
and
Brown-Wildenthal (2s1d shell) wave functions.  The magnitudes of
$\langle J_f \|O^{J=1}_{\rm {GT}} \| J_i \rangle$
were taken from measured $\beta$ decay $ft$ values.
\vskip12pt
\noindent
$^d$From Ref. [20].
\vskip12pt
\noindent
$^e$For this transition the calculated $\beta$ decay BGT is so small $(5 \cdot
10^{-5}$)
that theory cannot reliably determine the sign of the interference between
$O_{\rm {GT}}^{J=1}$ and
$O_{L = 2}^{J=1}$.  We have assumed constructed interference.

\pagebreak

\centerline{References}
\vskip18pt
\caption{[1]}{The GALLEX Collaboration, P. Anselmann et al.,
Phys. Lett. B 342 (1995) 440.}
\noindent
\vskip12pt
\caption{[2]}{The GALLEX Collaboration, P. Anselmann et al., Phys. Lett. B
285 (1992) 376, 314 (1993) 445, and 327 (1994) 377.}
\noindent
\vskip12pt
\caption{[3]}{See, for example, T. Kirsten, Nucl. Phys. B (Proc. Suppl.)
28 A (1992).}
\noindent
\vskip12pt
\caption{[4]}{J. N. Bahcall and R. K. Ulrich, Rev. Mod. Phys. 60 (1988)
297.}
\noindent
\vskip12pt
\caption{[5]}{J. N. Bahcall, Rev. Mod. Phys. 50 (1978) 881.}
\noindent
\vskip12pt
\caption{[6]}{M. Vergnes, talk presented at the Int. Conf. on the Structure of
 Medium-Heavy Nuclei, Rhodes, Greece (1979); M. Vergnes et al., Phys. Lett. B
72 (1978) 447.}
\noindent
\vskip12pt
\caption{[7]}{D. Ardouin et al., Phys. Rev C 18 (1978) 1201; G. Rotbard et
al., Phys. Rev. C 18 (1978) 86; S. Mordechai, H. T. Fortune, R.
Middleton, and G. Stephans, Phys. Rev. C 18 (1978) 2498.}
\noindent
\vskip12pt
\caption{[8]}{W. C. Haxton, in Nuclear Beta Decay and the Neutrino, ed. T.
Kotani, H. Ejiri, and E. Takasugi (World Scientific, Singapore, 1986), p.
225.}
\noindent
\vskip12pt
\caption{[9]}{G. J. Mathews, S. D. Bloom, G. M. Fuller, and J. N. Bahcall,
Phys.
Rev. C 32 (1985) 796.}
\noindent
\vskip12pt
\caption{[10]}{K. Grotz, H. V. Klapdor, and J. Metzinger, Astron. and
Astrophys.
154 (1986) L1.}
\noindent
\vskip12pt
\caption{[11]}{See, e.g., T. N. Taddeucci et al., Nucl. Phys. A 469
(1987) 125.}
\noindent
\vskip12pt
\caption{[12]}{D. Krofcheck et al., Phys. Rev. Lett. 55 (1985) 1051; D.
Krofcheck, Ph.D. thesis, Ohio State University (1987).}
\noindent
\vskip12pt
\caption{[13]}{Taken from the compilation of Ref. [15].}
\noindent
\vskip12pt
\caption{[14]}{S. M. Austin, N. Anantaraman, and W. G. Love, Phys. Rev. Lett.
73 (1994) 30.}
\noindent
\vskip12pt
\caption{[15]}{J. W. Watson et al., Phys. Rev. Lett. 55 (1985) 1369.}
\noindent
\vskip12pt
\caption{[16]}{S. Cohen and D. Kurath, Nucl. Phys. 73 (1965) 1.}
\noindent
\vskip12pt
\caption{[17]}{B. H. Wildenthal, Prog. Part. Nucl. Phys. 11 (1984) 5.}
\noindent
\vskip12pt
\caption{[18]}{The many-body matrix element of any one-body operator is
given exactly by
the one-body density matrix $\Psi_{\alpha \beta},$
\begin {eqnarray*}
\langle J_f\| O^{J=1} \|J_i\rangle = \sum_{\alpha \beta}
\Psi^{J=1}_{\alpha \beta} ~ \langle \alpha \| O^{J=1} \| \beta \rangle,
\end {eqnarray*}
where the sum extends over a complete set of single-particle quantum numbers
$\alpha$ and $\beta$.  The finite Hilbert spaces used in shell model
calculations approximate this expression by truncating the sums.  The
usual model space choice for A=71 is
2p$_{3/2}-$1f$_{5/2}-$2p$_{1/2}-$1g$_{9/2}$.
Regardless of the complexity of the many-body calculation within this space,
the matrix element of $O^{J=1}_{\rm {(p,n)}}$ can be written
\begin {eqnarray*}
& & - \Psi_{55} \sqrt{30 \over 7} (1 + {8 \over 5} \delta) + 2 \Psi_{33}
\sqrt{5 \over 3} (1 + {2 \over 5} \delta)
- {4 \over 3} \sqrt{3} \Psi^-_{31} (1 - {\delta \over 2}) \\
& & - \sqrt{2 \over 3} \Psi_{11} (1 + 4 \delta) + 6 \delta \sqrt{3 \over 5}
\Psi^-_{35}
+ \sqrt{110 \over 9} \Psi_{99} (1 + {8 \over 11} \delta),
\end {eqnarray*}
where $\Psi_{55} = \Psi_{1f_{5/2}1f_{5/2}}$, etc., and
$\Psi^-_{31} = \Psi_{31}-\Psi_{13}$, etc.
However, different model calculations may differ substantially in their
predictions for the $\Psi$s.}
\noindent
\vskip12pt
\caption{[19]}{In the limit where the initial $^{71}$Ga and final $^{71}$Ge
states are
simple 2p$_{3/2}$ proton and 1f$_{5/2}$ neutron holes in a closed $^{72}$Ge
core, respectively,
the only nonzero transition density matrix element is $\Psi_{35}$ = 1.  This
is the
maximum strength for states that can be written as simple Slater determinants.
  If one
introduces correlations, it is possible for $\Psi^-_{35}$ to exceed this
single-particle limit.  But, as the magnitude of $\Psi_{53}$ is expected to
be small,
we take the single-particle bound as an effective upper limit for
$\Psi^-_{35}$.}
\noindent
\vskip12pt
\caption{[20]}{B. Anderson, in Proc. GT and Neutrino Cross Section Workshop,
 ed. K. Lande (Univ. of Pennsylvania, unpublished), chapt. 13.}
\noindent
\vskip12pt
\caption{[21]}{One might be tempted to assume that the $^{71}$Ga normalization
is more accurate, since the (p,n) value for BGT(gs) agrees beautifully
with the $\beta$ decay result.  But, as we have seen that the $\beta$
decay/(p,n)
proportionality is unreliable for weak transitions, this
agreement is not a strong validation of the IAS normalization.}
\noindent
\vskip12pt
\caption{[22]}{J. N. Abdurashitov et al., Phys. Lett. B 328 (1994) 234; J.
Nico, talk presented at the Int. Conf. High Energy Physics, Glasgow, July,
1994.}
\noindent
\vskip12pt
\caption{[23]}{The Kamiokande Collaboration, K. S. Hirata
et al., Phys. Rev. D 38 (1988) 448 and 44
(1991) 2241; Phys. Rev. Lett. 65 (1990) 1297,
65 (1990) 1301, and 66 (1991) 9;
K. Nakamura, to be published in Proc. Int. Conf. on
Non-Accelerator Particle Physics (India) 1994.}
\noindent
\vskip12pt
\caption{[24]}{B. T. Cleveland et al., Nucl. Phys. B (Proc. Suppl.) 38 (1995)
47;
R. Davis, Jr., in Frontiers of
Neutrino Astrophysics, ed. Y. Suzuki and K. Nakamura (University Academy Press,
Tokyo, 1993) p. 47.}
\noindent
\vskip12pt
\caption{[25]}{These procedures follow N. Hata, S. Bludman, and P. Langacker,
Phys. Rev D. 49 (1994) 3622 and N. Hata and P. Langacker,
Univ. of Pennsylvania Report No. UPR-0625T (1994), to be published
in Phys. Rev. D.}
\noindent
\vskip12pt
\caption{[26]}{S. P. Mikheyev and A. Yu. Smirnov, Sov. J. Nucl. Phys. 42
(1985) 913 and Nuovo Cimento 9C (1986) 17; L. Wolfenstein, Phys.
Rev. D 17 (1978) 2369 and 20 (1979) 2634.}
\noindent
\vskip12pt
\caption{[27]}{J. N. Bahcall and M. H. Pinsonneault, submitted to Rev. Mod.
Phys. and Rev. Mod. Phys. 64 (1992) 885.}
\noindent
\vskip12pt
\caption{[28]}{S. Bludman, N. Hata, D. Kennedy, and P. Langacker, Phys.
Rev. D 47 (1993) 2220; N. Hata and P. Langacker, Phys. Rev. D
48 (1993) 2937 and D 50 (1994) 632.}
\noindent
\vskip12pt
\caption{[29]}{B. A. Brown and B. H. Wildenthal, Atomic Data and Nuclear Data
Tables 33 (1985) 347; W. T. Chou, E. K. Warburton, and B. A. Brown,
Phys. Rev. C 47 (1993) 163.}

\pagebreak

\centerline{Figure Captions}
\vskip18pt
\noindent
\caption{[1]}{Level scheme for $^{71}$Ge showing the excited states
that contribute to absorption of pp, $^7$Be, $^{51}$Cr, and $^8$B
neutrinos.}
\noindent
\vskip12pt
\caption{[2]}{Constraints imposed on BGT(5/2$^-$) and BGT(3/2$^-$) by
Krofcheck et al. [12] (small shaded region), by the present
reanalysis of the (p,n) results (area enclosed by the dashed lines),
and by the $^{51}$Cr source experiment (diagonal lines).}
\noindent
\vskip12pt
\caption{[3]}{The 90\% C.L. limits on $\phi(^7$Be) and $\phi(^8$B) imposed
by the SAGE and GALLEX experiments (a), by SAGE, GALLEX, and Kamiokande (b),
and by SAGE, GALLEX, Kamiokande, and $^{37}$Cl (c).
$\phi(^7\rm {Be})$ and $\phi(^8\rm {B})$ are in the SSM
units of 4.89E9/cm$^2$ s and 5.69E6/cm$^2$ s, respectively.  The allowed
regions are shaded.  The heavily shaded region corresponds
to the choice of maximum values for BGT(5/2$^-$) and BGT(3/2$^-$)
(so that excited state contributions to $^7$Be neutrino capture are
$\sim$ 90\% of the ground state contribution), while the entire
shaded region is allowed if no $^7$Be neutrino excited state capture occurs.
Thus the difference (the lightly shaded region) represents the effects of
nuclear structure
uncertainties on the extracted flux bounds prior to the source experiment.
The dotted line gives the result for Krofcheck et al. [12] BGT
values. Note that Fig. 3c also includes 99\% C.L. limits (left
unshaded for clarity).}
\noindent
\vskip12pt
\caption{[4]}{As in Figs. 3, but with the source experiment constraint
on BGT(5/2$^-$) and BGT(3/2$^-$) now included in the $\chi^2$ fit.
The residual unconstrained nuclear structure certainties are so
small that they are not shown (see text).  Also shown in (b) and (c) are
various standard and nonstandard solar model predictions (see Ref. [25] and
references therein).}
\noindent
\vskip12pt
\caption{[5]}{The MSW oscillation parameters allowed by the combined results
of the SAGE, GALLEX, Kamiokande, and $^{37}$Cl experiments, incorporating
the uncertainties in BGT(5/2$^-$) and BGT(3/2$^-$) determined by
the $^{51}$Cr source experiment.  We have employed
the fluxes from the BP SSM with He and metal diffusion [27].}

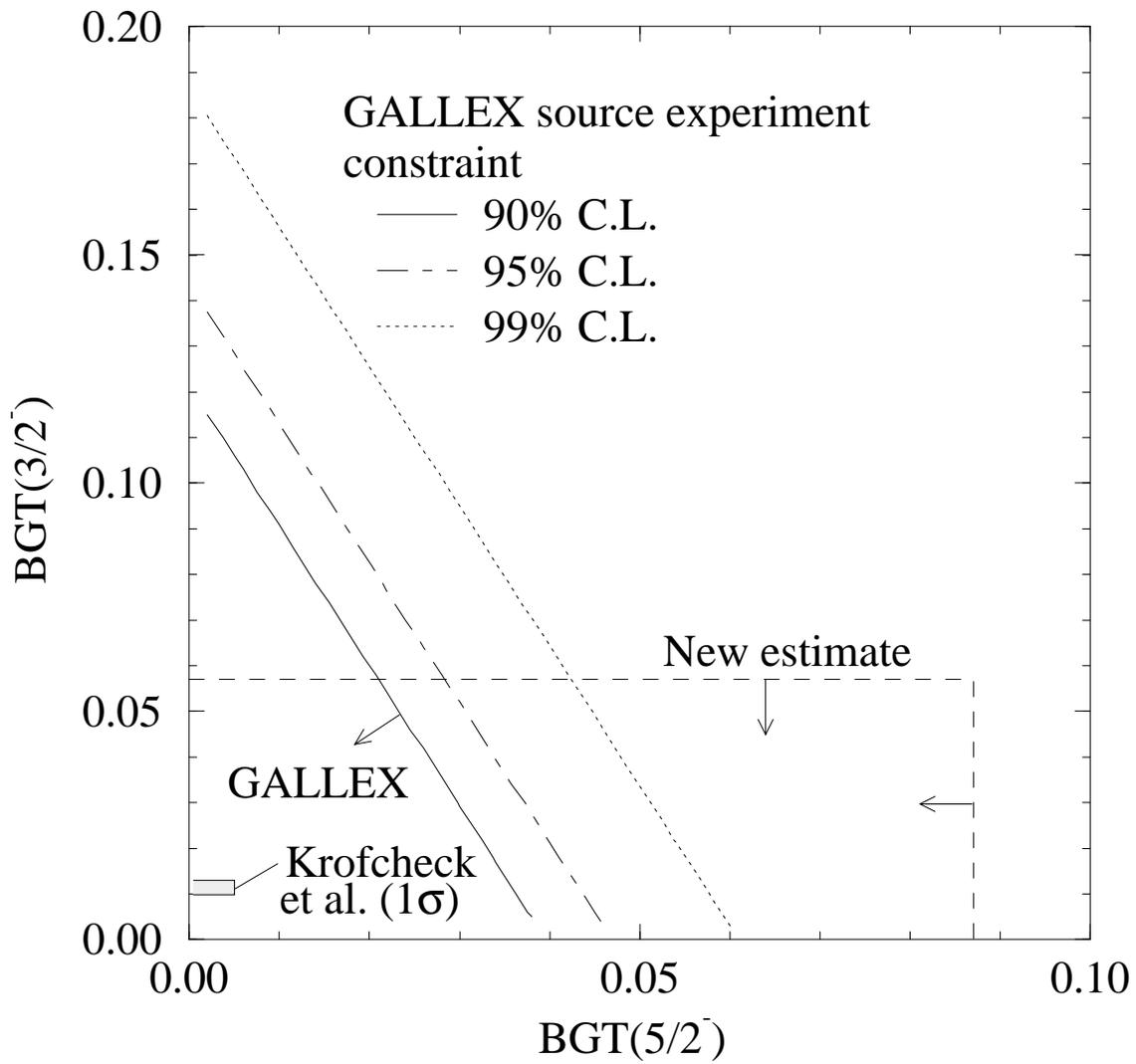
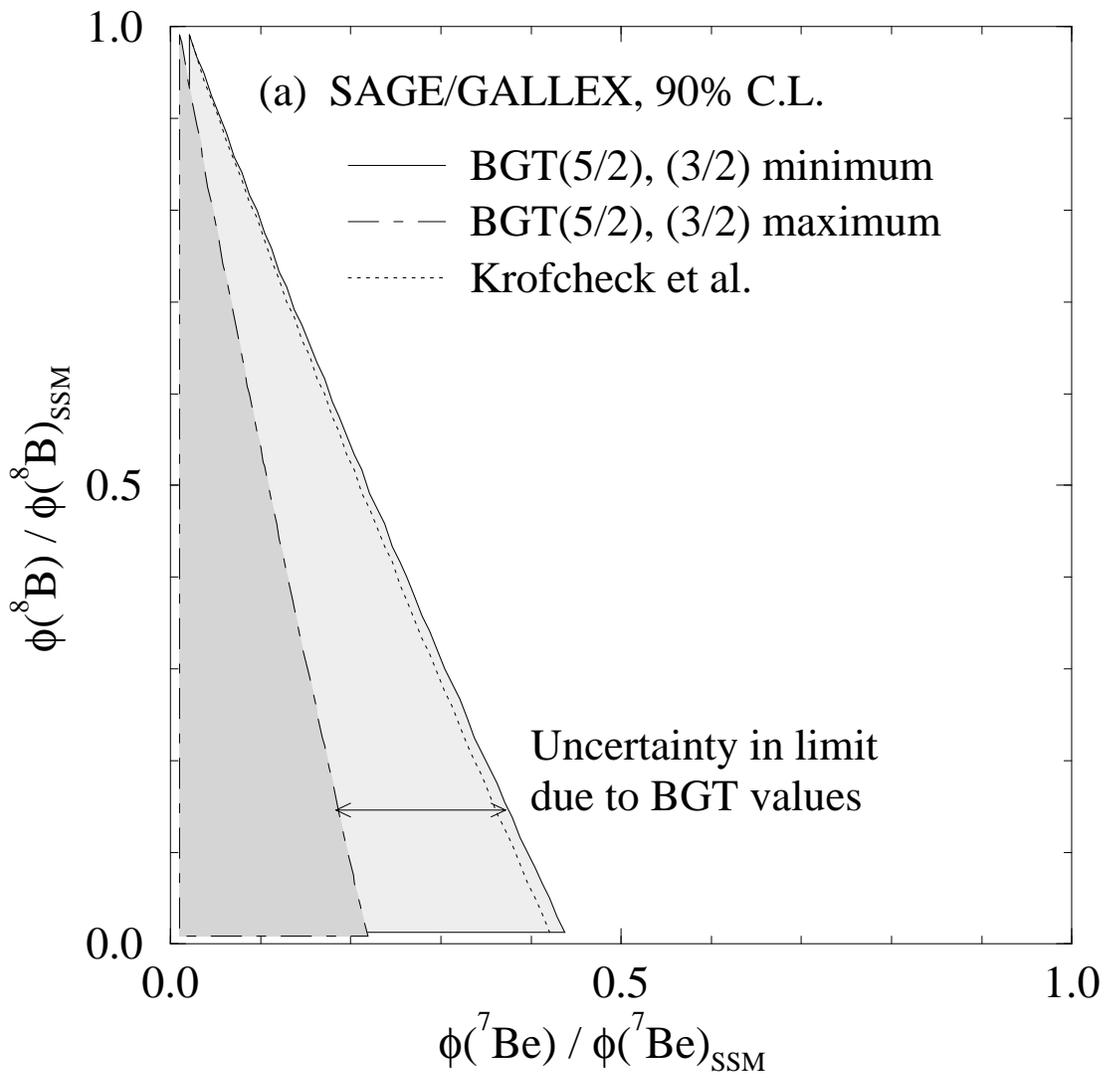
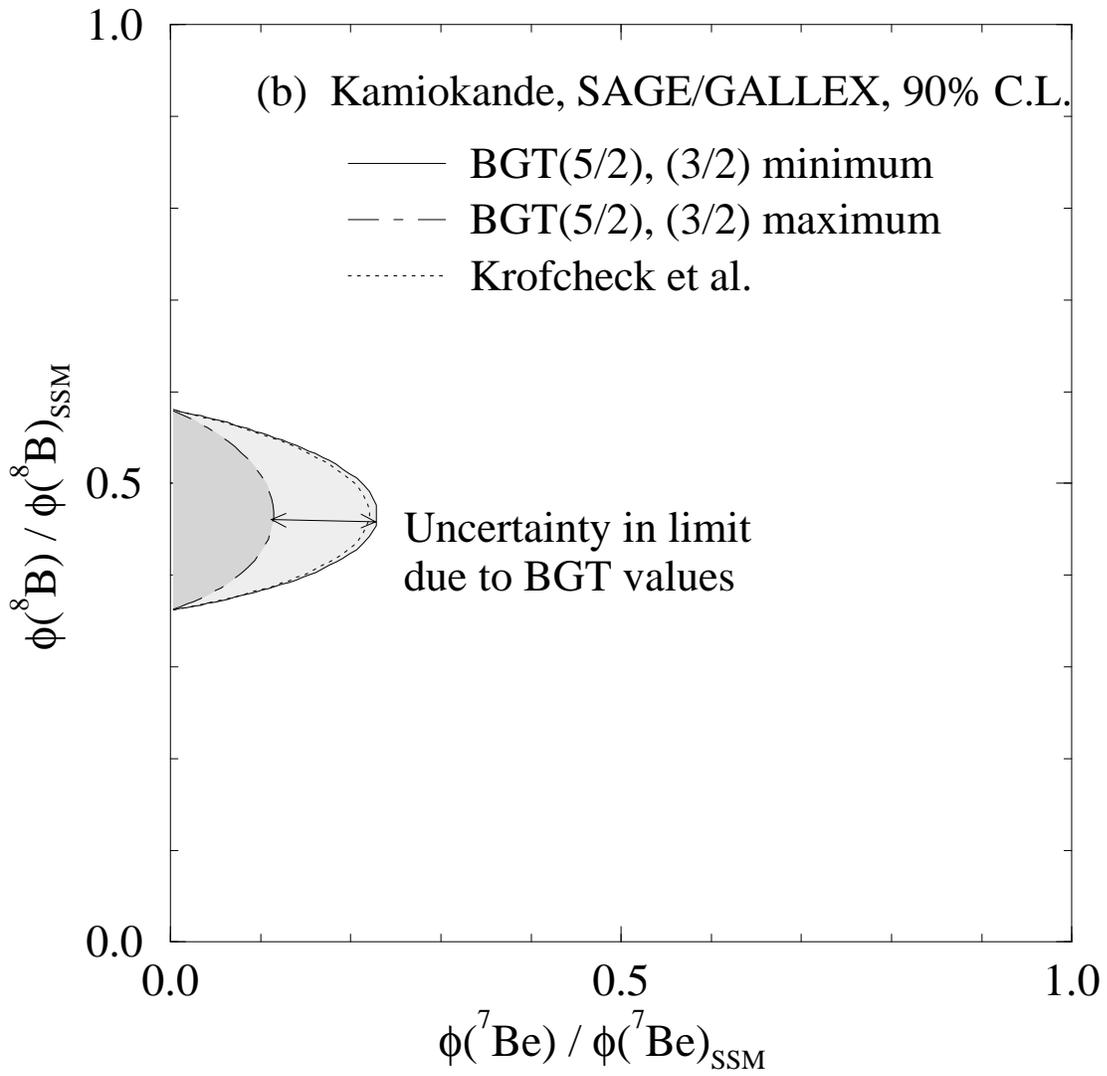
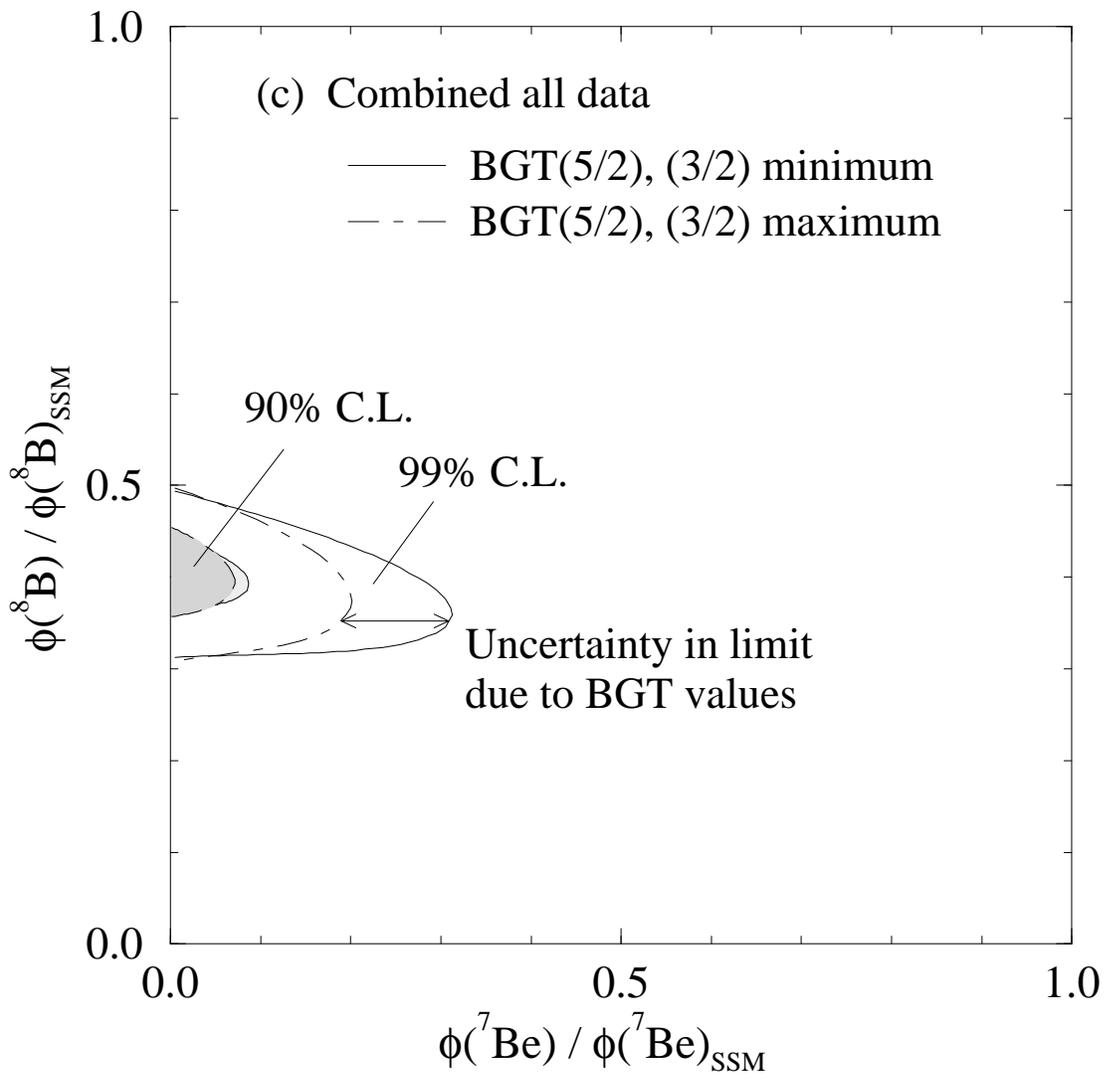
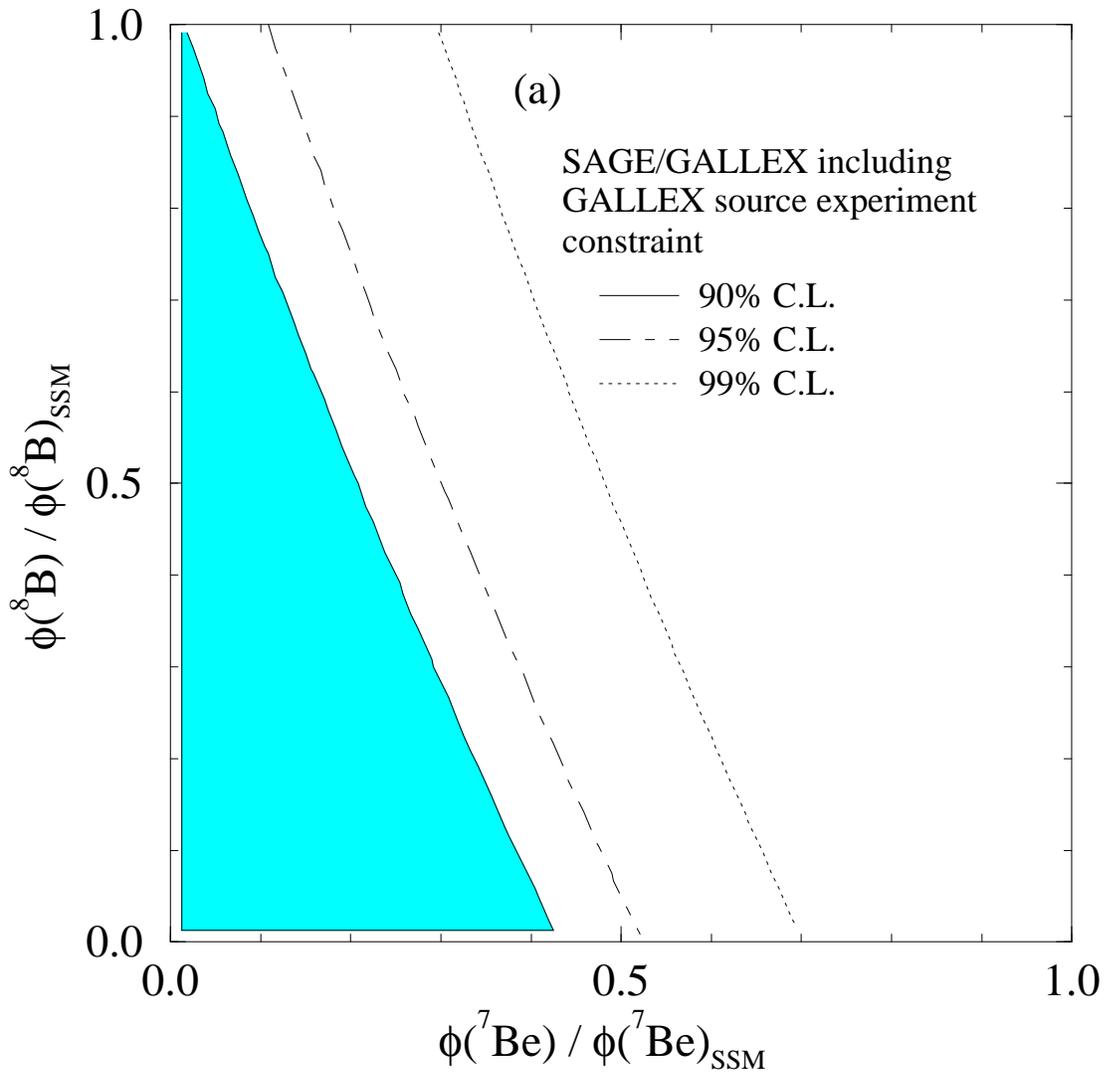
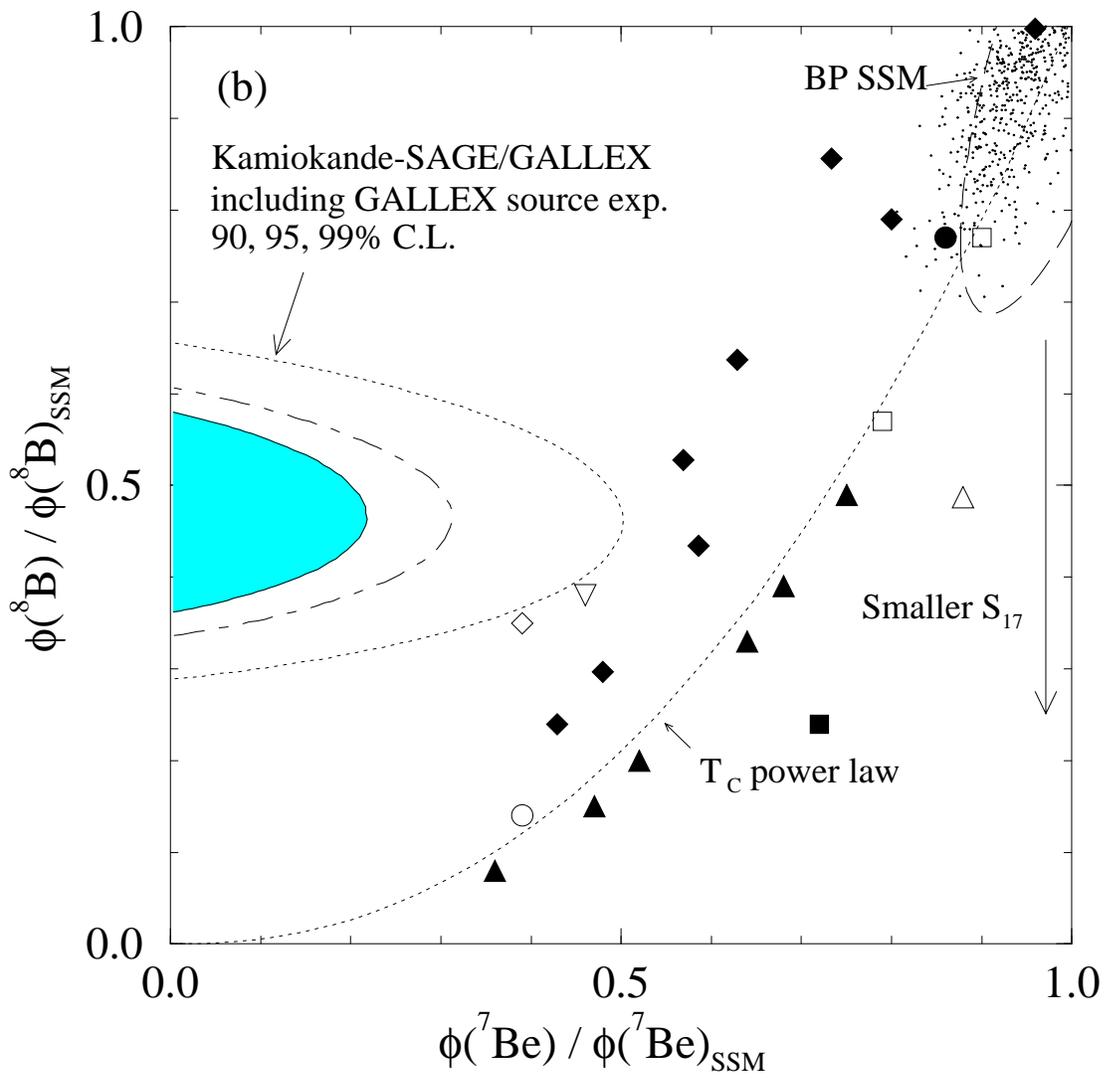
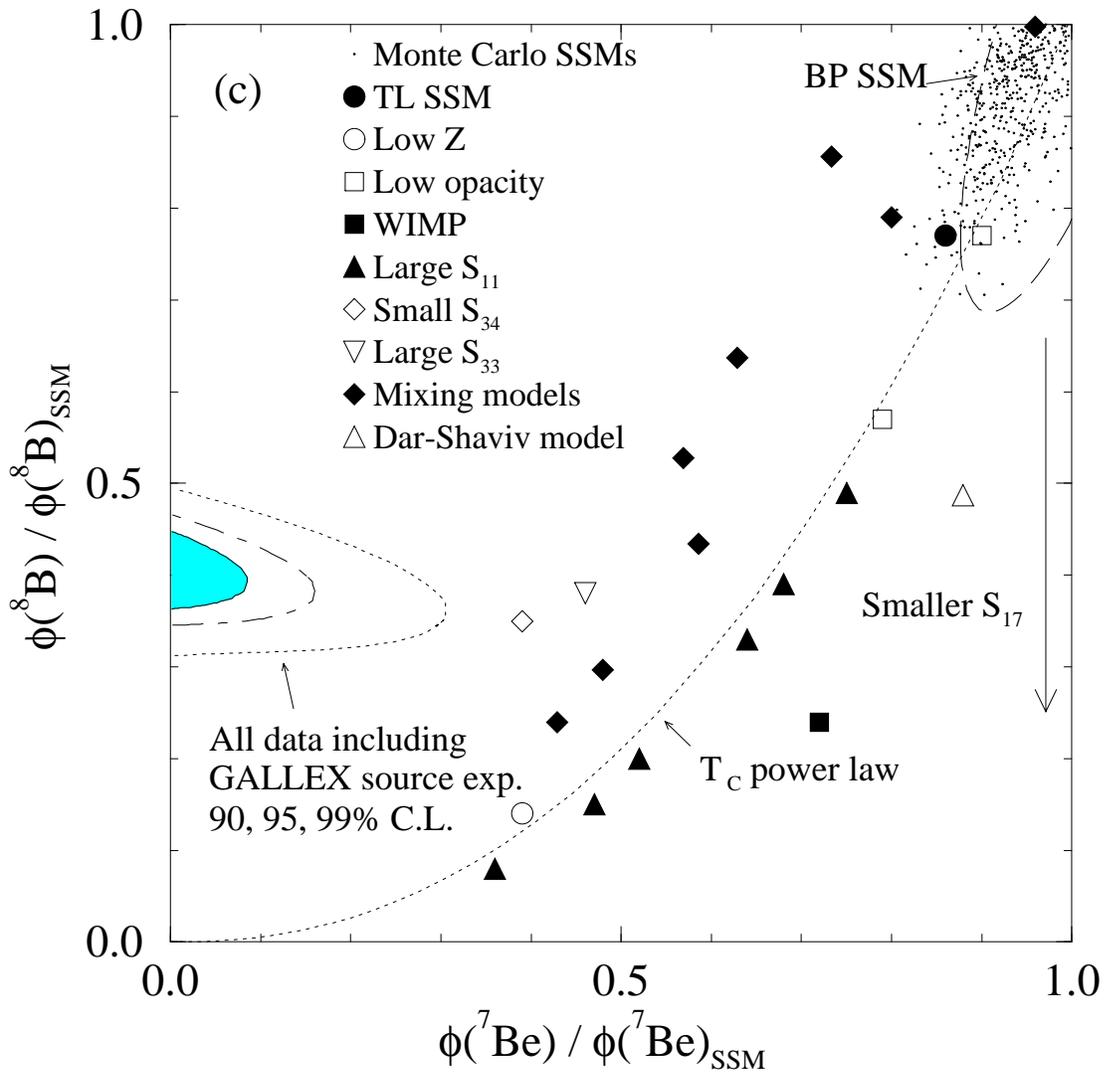
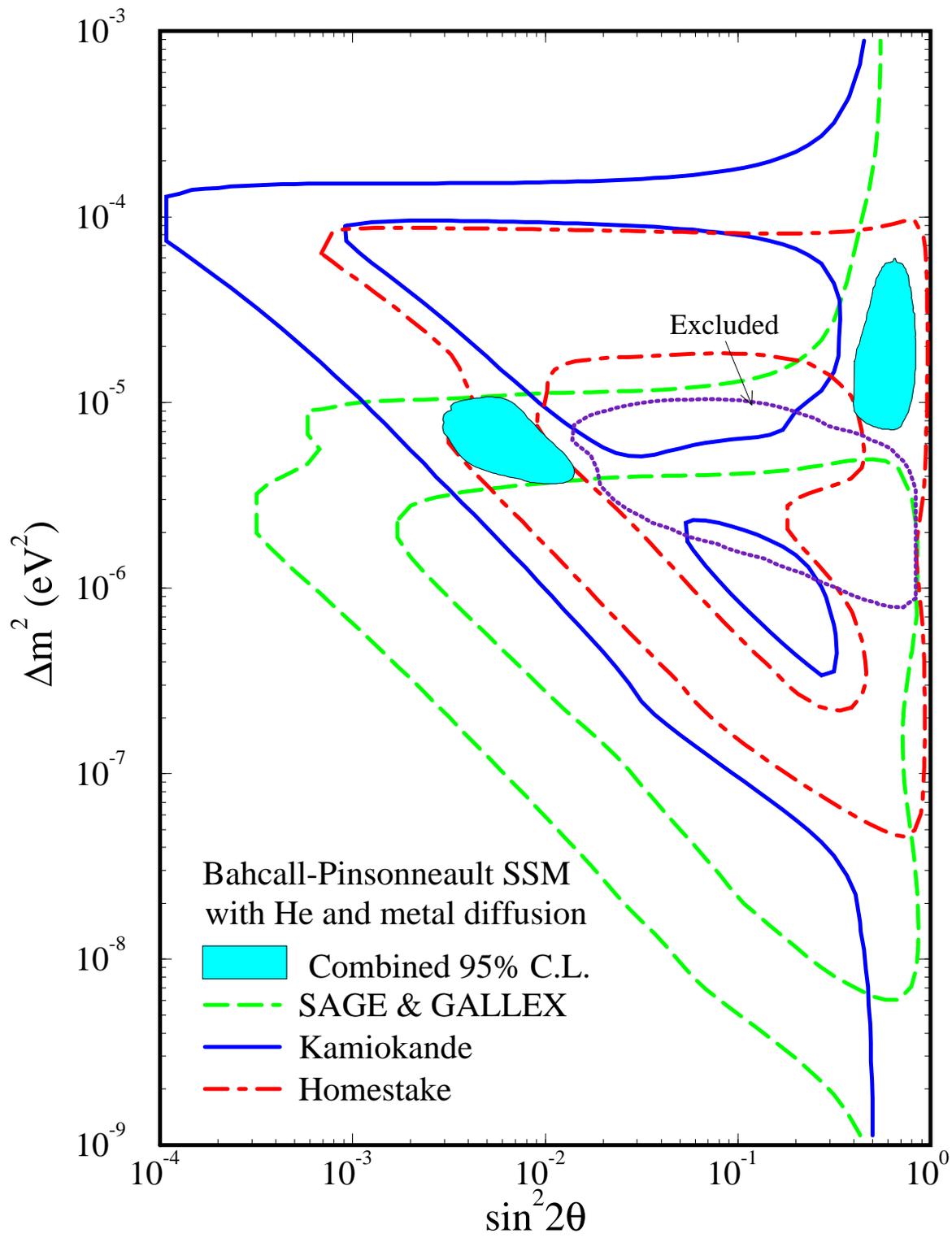
\end{document}